\documentclass[epj]{svjour}
\usepackage{graphics}
\begin{document}
\title{Electroweak Measurements of Neutron Densities in CREX and 
PREX at JLab, USA}
\author{C. J. Horowitz\inst{1}, K.S. Kumar\inst{2} \and R. Michaels\inst{3} 
}
\institute{Indiana University, Bloomington, Indiana, USA 
\and
University of Massachusetts, Amherst, Massachusetts, USA 
\and Thomas Jefferson National Accelerator Facility, Newport News, VA, USA}
\date{Received: date / Revised version: date}
%
\abstract{
Measurement of the parity-violating electron scattering
asymmetry is an established technique at Jefferson Lab and provides a
new opportunity to measure the weak charge distribution and hence pin
down the neutron radius in nuclei in a relatively clean and
model-independent way.  This is because the Z boson of the weak
interaction couples primarily to neutrons.  We will describe the
PREX and CREX experiments on ${}^{208}$Pb and ${}^{48}$Ca respectively;
these are both doubly-magic nuclei whose first excited state can be
discriminated by the high resolution spectrometers at JLab.
The heavier lead nucleus, with a neutron excess,
provides an interpretation of the neutron skin thickness in terms
of properties of bulk neutron matter.  For the lighter
${}^{48}$Ca nucleus, which is also rich in neutrons, microscopic
nuclear theory calculations are feasible and are sensitive to poorly
constrained 3-neutron forces.
} 
\authorrunning{Horowitz, Kumar, Michaels} 
\titlerunning{Neutron Densities from PREX and CREX}

\maketitle

\section{Introduction}
\label{intro}

Precise measurements of neutron densities provide a powerful probe of the symmetry energy $S$.  If $S$ increases with density, this will help move extra neutrons, in a neutron rich nucleus, from the high density interior into the low density surface region and create a neutron rich skin.  Therefore, measuring the thickness of this neutron skin allows one to infer the density dependence of the symmetry energy \cite{rocamaza1307}; see also the contribution to this volume by Vin${\tilde{\rm a}}$s {\it et al.} \cite{Vinas1308}.

The neutron skin thickness $\Delta R_{np}$ is the difference in r.m.s. neutron $R_n$ and proton $R_p$ radii,
\begin{equation}  
\Delta R_{np}= R_n- R_p\, .
\end{equation}
In light nuclei with $N \approx Z$, the neutrons and protons have similar density distributions. With increasing neutron number, the radius of the neutron density distribution becomes larger than that of the protons, reflecting the pressure from the symmetry energy. 

Proton radii have been determined accurately for many nuclei using electron scattering experiments \cite{chargeden,deVries,Angeli2013}. This accuracy reflects the accuracy of perturbative treatments of the electromagnetic process.  The neutron density distribution is more difficult to measure accurately because it interacts mainly with hadronic probes (pions \cite{pions1}, protons \cite{protons1,protons2,protons3}, antiprotons \cite{antiprotons1,antiprotons2}, and alphas \cite{ref:alpha1,ref:alpha2}) through nonperturbative interactions, the theoretical description of which is model-dependent.
Other approaches to inferring $R_n$ include inelastic scattering excitation of 
giant dipole resonances \cite{ref:GDR,PiekEpja}
and atomic mass fits \cite{Warda2009,Danielewicz2003}.
Neutron radii can also be measured with neutrino-nucleus elastic scattering \cite{1207.0693,PRD68.023005}.  Furthermore, new methods to detect low energy nuclear recoils will likely lead to important advances in neutrino scattering technology.  However, systematic errors may limit the precision of neutrino measurements of neutron radii.

Parity violation electron scattering, which arises from the weak interaction,
provides a theoretically clean method to measure neutron radii $R_n$.  This is because the weak charge of a neutron is much larger than that of a proton.  Therefore, the $Z^0$ boson, that mediates the weak neutral current, couples primarily to neutrons.  As a result, parity violation provides a model independent way to locate neutrons inside a nucleus, see section ~\ref{parity_violation}.  Parity violating experiments are difficult because the measured asymmetry $A_{PV}$
\begin{equation}
A_{PV} = \frac{\sigma_R - \sigma_L}{
\sigma_R + \sigma_L}
\label{eq:asy}
\end{equation}
where $\sigma_{R(L)}$ is the cross section for right (left)-handed helicity of the 
incident electrons, is very small, of order one part per million (ppm).  

Recently, the Lead Radius Experiment (PREX) at Jefferson Laboratory has pioneered parity violating measurements of neutron radii and demonstrated excellent control of systematic errors \cite{ref:prexI}.  The experimental configuration for PREX is similar to that used previously for studies of the weak form factor of the proton and $^4$He \cite{ref:happex}.  The Thomas Jefferson National Accelerator Facility provided excellent beam quality, while the large spectrometers in Hall A allowed PREX to separate elastically and inelastically scattered electrons and to greatly reduce backgrounds.           

This paper is organized as follows.  
In Section \ref{parity_violation} we describe the formalism 
for parity violating measurements of neutron densities, which
has been published in refs ~\cite{bigprex,ref:prexFF}.
The experimental methods are explained in Section \ref{experiment},
adapted from refs \cite{ref:happex,bigprex}
but containing some previously unpublished details.
The success of the methods was demonstrated by 
the PREX experiment on $^{208}$Pb, discussed in Section \ref{sec.prex}
based on  ~\cite{ref:prexI}.
The main new material is the discussion of a planned 
follow-on measurement PREX-II \cite{ref:prexII} in Section \ref{sec.prex}
and the CREX proposal for $^{48}$Ca \cite{ref:CREX} in Section \ref{CREX}, 
with an outlook based on the recent CREX workshop \cite{CREXworkshop}.
These experiments PREX-II and CREX should measure 
neutron skins with high accuracy.  
We conclude in Section \ref{conclusions}. 

\section{Parity-Violating Measurements of Neutron Densities}
\label{parity_violation}

In the Born approximation, the parity violating cross-section asymmetry for
longitudinally polarized electrons elastically scattered 
from an unpolarized nucleus,  $A_{PV}$, is
\begin{equation}
A_{PV}\approx \frac{G_FQ^2}{4\pi\alpha\sqrt{2}}
\frac{F_W(Q^2)}{F_{ch}(Q^2)}
\label{born_asy}
\end{equation}
where  $G_F$ is the Fermi constant, $\alpha$ the
fine structure constant, and $F_{ch}(Q^2)$ is the Fourier 
transform of the known charge density.  The asymmetry is proportional to the weak form factor $F_W(Q^2)$.  This is closely related to the Fourier transform of the neutron density (see below), and therefore the neutron density can be extracted from an electro-weak measurement \cite{dds}.

However, the Born
approximation is not valid for a heavy nucleus and Coulomb-distortion effects
must be included.  These have been accurately calculated \cite{couldist}
because the charge density is well known, and many
other details relevant for a practical parity-violation experiment to 
measure neutron densities have been 
discussed in a previous publication~\cite{bigprex}.    

The weak form factor is the Fourier transform of the weak charge density $\rho_W(r)$,
\begin{equation}  
F_W(Q^2)=\frac{1}{Q_W}\int d^3r \frac{\sin Q r}{Q r} \rho_W(r),
\label{F(q)}
\end{equation}
and is normalized $F(Q=0)=1$.  The total weak charge of the nucleus is $Q_W=\int d^3r \rho_W(r)$.
For $\rho_W(r)$ of a spin zero nucleus, we neglect meson exchange and spin-orbit currents and write \cite{bigprex}
\begin{equation}
\rho_W(r)=4\int d^3r'\bigl[G_n^Z(|{\bf r}-{\bf r'}|)\rho_n(r')+G_p^Z(|{\bf r}-{\bf r'}|)\rho_p(r')\bigr]\, .
\label{rhoWinitial}
\end{equation}
Here the density of weak charge in a single proton $G_p^Z(r)$ or neutron $G_n^Z(r)$ is the Fourier transform of the nucleon (Electric) Sachs form factors $G_p^Z(Q^2)$ and $G_n^Z(Q^2)$.  These describe the coupling of a $Z^0$ boson to a proton or neutron \cite{bigprex},
\begin{equation}
4G_p^Z=q_pG_E^p+q_n G_E^n -G_E^s,
\end{equation}
\begin{equation}
4G_n^Z=q_nG_E^p + q_p G_E^n -G_E^s.
\end{equation}
At tree level, the weak nucleon charges are $q^0_n=-1$ and $q_p^0=1-4\sin^2\Theta_W$.  We include radiative corrections by using the values $q_n=-0.9878$ and $q_p=0.0721$ based on the up $C_{1u}$ and down $C_{1d}$ quark weak charges in ref  \cite{erleretal,stdmodelpdb}. The Fourier transform of the proton (neutron) electric form factor is $G_E^p(r)$ ($G_E^n(r)$) and has total charge  $\int d^3r G_E^p(r)=1$  ($\int d^3r G_E^n(r)=0$).  Finally $G_E^s$ describes strange quark contributions to the nucleon's electric form factor \cite{SAMPLEsff,G0sff,A4sff,ref:happex}.  

Given that $q_p$ is small, and strange quark contributions have been greatly limited by previous measurements, we see that $\rho_W(r)$ is primarily the point neutron density $\rho_n(r)$ folded with the weak form factor of a single neutron.  The neutron density is normalized to the number of neutrons $\int d^3r \rho_n(r)=N$, while the proton density is normalized to the number of protons $\int d^3r \rho_p(r)=Z$.  The point neutron radius $R_n$ is defined from $R_n^2 = \int d^3r r^2 \rho_n(r)/N$.

Measuring $A_{PV}$ for a single low $Q^2$ allows one to infer $F_W(Q^2)$ and from it the r.m.s. weak radius $R_W$.  This is then related to $R_n$ \cite{ref:prexFF}
\begin{equation}
R_n^2=\frac{Q_W}{q_nN}R_W^2 -\frac{q_p Z}{q_n N} R_{ch}^2 -\langle r_p^2 \rangle -\frac{Z}{N}\langle r_n^2\rangle +\frac{Z+N}{q_nN}\langle r_s^2\rangle\, .
\label{RNsingleQ}
\end{equation}
Here the (known) charge radius of the nucleus is $R_{ch}$, the square of the charge radius of a single proton is $\langle r_p^2\rangle=0.769$ fm$^2$ and that of a single neutron is $\langle r_n^2\rangle=-0.116$ fm$^2$, finally $\langle r_s^2\rangle=\int d^3r' r'^2 G_E^s(r')$ is the square of the nucleon strangeness radius.  Previous measurements \cite{ref:happex,SAMPLEsff,G0sff,A4sff}
have constrained this to be small.  
Note that the $-\langle
r_p^2\rangle$ term in eq ~\ref{RNsingleQ} comes from the weak radius 
of a single neutron.  This is related to the charge radius of a single proton.

To summarize this section, measuring $A_{PV}$ determines the weak form factor $F_W(Q^2)$ and from this the neutron radius $R_n$.  The neutron skin thickness $R_n-R_p$ then follows, since $R_p$ is known.  Finally, the neutron skin thickness constrains the density dependence of the symmetry energy.

\section{Experimental Method}
\label{experiment}

\subsection{Overview of the Method}
\label{method_overview}

The experiments run at Jefferson Lab using the high-resolution
spectrometers (HRS)~\cite{Alcorn:2004sb}
in Hall A, comprising a pair of 3.7 msr spectrometer 
systems with $10^{-4}$ momentum resolution,
which focus elastically scattered electrons onto 
total-absorption detectors in their focal planes. The ``hardware
momentum resolution" of the spectrometer system i.e. the width of
the distribution for mono-energetic electrons with no event-by-event corrections,
is better than $10^{-3}$, so that the elastic electrons populate a region that
is otherwise free from contamination from inelastic events.

A polarized electron beam scatters from a target foil,
and ratios of detected flux to beam current
integrated in the helicity period are formed (so-called ``flux integration"),
and the parity--violating asymmetry in these
ratios computed from the helicity--correlated
difference divided by the sum (eq ~\ref{eq:asy}).
Separate studies at lower rates are required to
measure backgrounds, acceptance, and $Q^2$.  
Polarization is measured once a day by a
M{\o}ller polarimeter, and monitored continuously
with the Compton polarimeter.

The asymmetry is small, of the order of
one or two parts per million (ppm) for the kinematics of interest
for the two nuclei under primary consideration namely, $^{208}$Pb (PREX) and $^{48}$Ca (CREX). 
To have significant impact on our knowledge of skin thicknesses, $A_{PV}$
must be measured with a precision in the range of 3\% or better 
(see fig \ref{fig:R}).
Experiments of this
nature are optimized to the challenges of precision measurement 
of very small asymmetries,
which require high count rates and low noise to achieve
statistical precision as well as a
careful regard for potential systematic errors
associated with helicity
reversal, which must be maintained below the $10^{-8}$ level
(see section ~\ref{systematics}).

One common feature of all measurements of parity-violation in 
electron scattering is a rapid
flipping of the electron beam helicity, allowing a differential
measurement between opposing
polarization states on a short timescale. The enabling technology 
for these measurements lies
in the semiconductor photo-emission polarized electron source, 
which allows rapid reversal of
the electron polarization while providing high luminosity, 
high polarization, and a high degree
of uniformity between the two beam helicity states.
Developments with the polarized source at
Jefferson Lab are critical to the success 
of this program~\cite{Sinclair:2007ez}.

\par

In a parity experiment,
the asymmetry generally increases with $Q^2$ while the cross
section decreases, which leads to an optimum choice of kinematics.
For parity-violating neutron density experiments, 
the optimum kinematics 
is the point which effectively
minimizes the error in the neutron radius $R_n$.  
This is equivalent to maximizing 
the following product, which is the figure-of-merit (FOM)   

\begin{equation}
FOM =   R \times A^2 \times {\epsilon}^2
\label{eq:FOM}
\end{equation}

Here, $R$ is the scattering rate,
$A$ is the asymmetry, $\epsilon = \frac{dA/A}{dR_n/R_n}$ is the 
the sensitivity of the asymmetry for a small change in $R_n$, 
$dR_n/R_n$ is a fractional change in $R_n$ and $dA/A$ is a 
corresponding fractional change in $A$.  Note that the FOM
defined for many types of parity-violation experiments is $R \times A^2$,
but the neutron-density measurements must also fold in the
sensitivity $\epsilon$.

Given practical constraints on the solid angle of the
HRS, the optimization algorithm favors smaller scattering angles.
Using septum magnets we reach $\sim 5^{\circ}$ scattering angle.
Once the angle is fixed, the optimum energy for elastic scattering can be specified.
Simulations that are performed to design the experiment include 
the Coulomb distortions, as well as radiative losses, 
multiple scattering, and ionization losses in materials, 
together with a model for the tracking of particle 
trajectories through the HRS and septum magnets.

\par The two nuclei of interest for 1\%, or better, $R_n$ measurements 
($^{48}$Ca and $^{208}$Pb) are equally accessible experimentally and have been
very well studied \cite{chargeden,wise,cavendon,emrich,quint}.
These are doubly-magic and have a simple nuclear structure, making
them good candidates for extracting the symmetry energy.
Each nucleus has the advantage that it has a large 
splitting to the first excited state (2.60 MeV for $^{208}$Pb
and 3.84 MeV for $^{48}$Ca), thus lending themselves 
well to the use of a flux integration technique.  

\subsection{Control of Random and Systematic Fluctuations}
\label{systematics}

To achieve the $10^{-8}$ statistical precision and systematic control 
for $A_{PV}$ measurements requires a precise
control and evaluation of systematic errors, as
has been developed at Jefferson Lab~\cite{ref:happex} 
and elsewhere~\cite{Humensky:2002uv}.
The apparatus must have the ability of measuring rates in excess of 1 GHz with negligible deadtime.
In this section we will discuss some of the 
details of the techniques involved.


The polarized electron beam originates from a strained GaAsP 
photocathode illuminated by 
circularly polarized light \cite{Sinclair:2007ez}.
Several monitoring devices measure the
beam's intensity, energy, polarization.
The sign of the laser circular polarization determines
the electron helicity; this 
is held constant for periods of typically 8 ms, 
referred to as ``windows''.  
The integrated responses of detector PMTs and beam monitors
are digitized by an 18-bit ADC and recorded for each window.
The helicity states are arranged in
patterns, for example ($+--+$ or $-++-$) for a quadruplet
structure with each window 8.23 msec long.
These patterns ensure that
complementary measurements are made at the same phase 
relative to the 60~Hz line power, 
thus canceling power-line noise from the asymmetry measurement.

The signals are integrated over the helicity window
because the rates are too high for a counting DAQ.
The right-left helicity asymmetry in the integrated detector response,
normalized to the beam intensity,
is computed for sets of complementary helicity windows in each
quadruplet to form the raw asymmetry
$A_{raw}$.   The sequence of these patterns is
chosen with a pseudo-random number generator.  
The reversals of the beam helicity occur in a random sequence in
order to uncouple them from other
parameters which affect the cross section.
To take full advantage of the high scattered flux and to ensure that $A_{raw}$ measurement
fluctuations are dominated by counting statistics, the electronics chain is designed to
be capable of measuring the response of each helicity window with a precision better than $10^{-4}$.

The requirement of high statistics also requires high current on a relatively thick target. For the case of
PREX, an isotopically pure $^{208}$Pb 0.55 mm thick target is used. 
Two 150~$\mu$m diamond foils sandwich the lead foil to 
improve thermal conductance to a copper 
frame cooled to 20K with cryogenic helium.  
Non-uniformities in target thickness due to thermal damage could cause 
window-to-window luminosity fluctuations from variations in the target 
area sampled by the rastered beam. This potential source of random noise is
controlled by locking the raster
pattern frequency to a multiple of the helicity frequency.  
Low-current calibration data, triggered
on individual scattered electrons, are regularly collected to 
evaluate the thickness
of lead relative to diamond.

The sensitivity of the
cross section to fluctuations in the beam parameters, as well as the
helicity correlated differences in them must be accurately monitored 
concomitant with the collection of physics data.
Care must be taken to isolate the helicity signals, since
electronic pickup of the helicity correlated signals
could cause a false asymmetry.
For an integrating DAQ system, the linearity
of the detector electronics and the susceptibility to 
backgrounds are important issues.
The particle detectors for the scattered electons
are quartz \footnote[1]{Artificial fused silica, brand name 
Spectrosil 2000 from Quartz Plus, Inc} bars.
Cherenkov light from each quartz bar traverse air  
light guides and are detected by photo-multipliers (PMT). 

While a quartz Cerenkov detector is valued for radiation 
hardness and insensitivity to soft backgrounds, 
there is a particular challenge for electrons with energy less than 2 GeV.  
In this energy range, shower fluctuations in a thick or radiated
detector significantly degrade energy resolution, 
while photon statistics degrade the 
energy resolution for a thin detector.  
The energy resolution $\Delta E$ at nominal
electron energy $E$ increases the statistical error that one would
have with infinite resolution $\sigma_0$ to obtain the total
statistical error $\sigma = \sigma_0 \sqrt{1 + {(\frac{\Delta E}{E})}^2}$.
During PREX-I, the detector thickness was optimized and achieved sufficient
energy resolution so that the statistical degradation factor was 1.06. 

To study and help cancel the helicity correlated systematics,
there should be more than one way to change the sign of
the beam helicity.
A half-wave ($\lambda$/2) plate was periodically inserted into the 
injector laser optical path, reversing the
sign of the electron beam polarization relative to both the 
electronic helicity control signals
and the voltage applied to the polarized source laser electro-optics.
Roughly equal statistics were collected with this waveplate 
inserted and retracted, suppressing 
many possible sources of systematic error.  

An independent method of 
helicity reversal was
feasible with a pair of Wien spin-rotators separated by a solenoid, 
 providing an additional powerful
check of systematic control.  
Reversing the direction of the solenoidal field reversed the electron 
beam helicity 
while the beam optics, 
which depend on the square of the solenoidal magnetic field, were unchanged.  
The $\lambda/2$ reversal was done about every 12 hours and the 
magnetic spin reversal was performed every few days. The dataset 
consisting of a period between two successive 
$\lambda/2$ or magnetic spin-reversals is referred to as a ``slug''.

\subsection{Beam Induced Asymmetries}

\begin{figure}
\resizebox{0.48\textwidth}{!}{%
\includegraphics{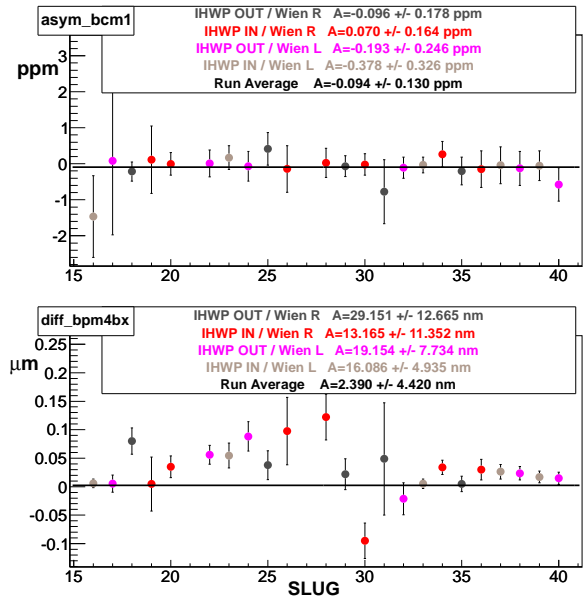}
}
\caption{
PREX-I helicity-correlated charge asymmetries (top) and position 
differences (bottom) on a representative monitor versus slug 
(a slug is $\sim 1$ day of running).  The different colors correspond
to four different combinations of IHWP and Wien used for slow sign reversal, as 
explained in the text.  
To illustrate the systematics, the data points are plotted without sign correction for the helicity flip.  
The final average with 
all sign corrections is shown by the black horizontal bar and was controlled at 
the 5 nm level averaged over the PREX-I run.  The charge asymmetry was forced 
to zero by the standard feedback system.}
\label{beam_corr}
\end{figure}

PREX-I was able to achieve overall
asymmetry corrections due to helicity-correlated beam position fluctuations of 
about $40~\mathrm{ppb}$ with position differences $<5~\mathrm{nm}$.  The position/asymmetry
correlations are measured using two independent methods: 
first, directly observing the asymmetry correlations by the natural beam motion and second, by
systematically perturbing the beam through a set of magnetic coils (dithering).
Achieving these small values for the differences was possible in 
part by periodically inserting the half-wave
plate in the injector and flipping the helicity of the beam using a double-Wien
filter which helps them cancel over time.
Fig ~\ref{beam_corr} shows the helicity-correlated charge asymmetries and 
position differences versus time during PREX-I.  A beam current monitor
(BCM) and one representative 
beam position monitor (BPM) is shown; the other BPMs
look similar. Feedback on the charge asymmetry forced it to be zero
within 0.13 ppm.  The utility of the
slow reversals is demonstrated the BPM difference plot; without them, the 
position differences remained at the $\sim$ 50 nm level (the points without
sign correction) averaged over the experiment; with the reversals,
the differences averaged to the $\sim 5$ nm level (the black lines)
and became a negligible 
correction \cite{KiadThesis,RupeshThesis,LuisThesis,MindyThesis,ZafarThesis}.

The correction made for PREX-I was dominated by fluctuations in 
the scattered beam
intensity due to small changes in the accepted angle and the sharply falling
lead cross section.  CREX will run at a 
higher $Q^2$ ($0.022~\mathrm{(GeV/c)}^2$), and since ${}^{48}$Ca is a smaller
nucleus, $\frac{d\sigma}{d\theta}/\sigma$ is smaller by a factor of 4.3.
We conservatively estimate that the uncertainty on the corrections 
for CREX will be $\sim 7~\mathrm{ppb}$, the same as for PREX-I.

The integrated signals in the helicity windows are normalized to the beam
current monitor signals to remove helicity correlated beam intensity fluctuations.
Non-linearities in the BCMs produce additional false asymmetries, which are related
to the overall charge asymmetry.  Based on past running, we can expect an accumulated
charge asymmetry less than $100~\mathrm{ppb}$ and an uncertainty on that 
correction of 1.5\%, so $1.5~\mathrm{ppb}$, or 0.1\% propagated to the final asymmetry.

\subsection{Contributions from Inelastic States and Isotopes}
\label{inelastics}

While doubly-magic nuclei are preferred for 
their simple theoretical structure, they are 
also preferred experimentally because they 
have a large energy separation between 
elastic scattering and the first excited levels, 
as mentioned in section ~\ref{method_overview}.
For  $^{208}$Pb (${}^{48}$Ca) the separation is 2.60 (3.84) MeV.
Using the HRS spectrometers with $10^{-3}$ hardware momentum resolution,
we can place the elastic peak on our detectors, while ensuring
that the inelastic electrons are not intercepted.

In ref \cite{bigprex} the asymmetry for 
the 2.60 MeV \hskip 0.04in $3^-$ state of 
 $^{208}$Pb was calculated using a model of collective nuclear excitations,
assumed to be isoscaler, 
and found to be comparable to the elastic asymmetry, with the result
$A(3^-) \approx 1.25 A_{\rm elastic}$ and a relative uncertainty of 35\%
mainly due to unknown Coulomb distortions.  In PREX, the 
relative rates of inelastic electrons were measured 
with a thin lead target in counting mode \cite{KiadThesis}.
The first state was 
at the edge of our detector acceptance, and the rate multiplied by the
acceptance was $< 4\times10^{-4}$.  Higher excited states were even
further away from our detector and their rates 
were negligible \cite{KiadThesis}.

The lead is 99.1\% pure $^{208}$Pb with 0.7\% of $^{207}$Pb,
0.2\% of $^{206}$Pb, and negligible
amounts ($< 10^{-4}$) of other elements.  
The ${}^{48}$Ca 
target is 99\% chemically pure and the only important isotope
contamination is a 3\% contamination from $^{40}$Ca.
We have evaluated the amount of inelastic ${}^{48}$Ca contamination 
based on form factor measurements 
of electron scattering ~\cite{wise} which 
covered the same momentum transfer range of CREX.
Elastic and inelastic events were simulated using our transport model
for the HRS with the septum magnet.
The first excited state at 3.84 MeV has a cross section that is 
0.94\% of the elastic cross section, and the placement of the detector 
suppresses this to a 0.19\% background.  The next most 
important contribution is the second excited state at 4.51 MeV,
contributing 0.18\% background.  Altogether, the 
first ten inelastic states of $^{48}$Ca produce a 0.4\% background.
This might be further reduced with fine-tuning of the spectrometer
optics and detector geometry.  The ground state for the $^{40}$Ca 
isotope contaminant is 
$J^\pi = 0^+$ and the asymmetry should be reliably calculable; 
the first excited state
at 3.4 MeV is also $0^+$ and is mostly suppressed by the HRS.

Based on refs \cite{Walecka77,donnelly1,donnelly2,donnelly3},
a simple order-of-magnitude
estimate for the e$-$Nucleus asymmetry can be obtained at forward
angle where the axial-hadronic contributions to the asymmetry are
suppressed both kinematically as well as by the 
smallness of $[1 - 4{\rm sin}^2 \theta_W]$.

\begin{equation}
A \hskip 0.04in \approx \hskip 0.02in +(-) \hskip 0.05in
\frac{G_F Q^2}{4 \pi \alpha \sqrt 2}
\label{asymestimate}
\end{equation}

Here ''$+$'' is for an isoscaler transition and ``$-$'' is
for isovector.
We've made the following cross-checks of eq \ref{asymestimate}.
We find agreement within a few percent with the Feinberg's formula \cite{Feinberg}
for the asymmetry for elastic scattering
from $^{12}$C, and a similarly good agreement
at forward angle with the asymmetry from the lowest isovector 
excited state of $^{12}$C at 15.1 MeV 
calculated in refs \cite{donnelly1,oconnell}.
Next, eq \ref{asymestimate} agrees within a factor of 2
with Horowitz's result \cite{bigprex} for
the $3^-$ state of $^{208}$Pb, mentioned above,
as well as if that state were assumed to be isovector.

We believe the low-lying states, as well as elastic scattering, should be
be predominantly isoscaler.  
Therefore, the asymmetries are not expected to be significantly 
different from the measured asymmetry.
More accurate calculations of the asymmetry for the isotopes
and for low-lying states of ${}^{48}$Ca would be helpful.
The inelastic contamination will also be measured during the experiment
using the standard detectors in counting mode.

\subsection{$Q^2$ Measurement}         
\label{QsqMeas}

A measurement of $Q^2$ to better than 1\% is needed in order
to interpret the asymmetry and extract neutron densities,
because the asymmetry is a strong function of $Q^2$.
For example, for $^{208}$Pb  the sensitivity is 
$\mathrm{d}A_{PV}/\mathrm{d}Q^2 \approx 30~\mathrm{ppm}/\mathrm{GeV}^2$
at the kinematics of PREX.

Measuring the small scattering angle is the primary challenge.
Survey techniques, while being a good cross-check, are
insufficient to constrain the propagated uncertainty to less than 1\%.
A nuclear recoil technique using a water
cell target~\cite{ref:prexI,ref:happex,KiadThesis,QsqReports} 
limits the scale error 
on $\langle Q^2 \rangle$ to $1\%$.
By comparing the energy difference between the elastically
scattered electrons from the protons in water to the elastic
peak from $^{16}$O and other heavy nuclei in the water target,
the absolute angle can be fixed.
This technique was used for PREX and obtained an absolute
angle determination of about $0.4~\mathrm{mrad}$ \cite{KiadThesis}.
Anticipating a comparable energy resolution and the kinematic
differences to CREX, an angular determination of
$0.28~\mathrm{mrad}$ is expected.

\subsection{Asymmetry Analysis}
\label{data_analysis}

The result of analysis is an asymmetry at a particular 
kinematic point.  The data used
to compute the asymmetry must pass loose requirements on beam quality,
but no helicity-dependent cuts are applied.  
The integrated response for the detectors, $D$, and for each 
beam monitor is digitized and recorded for each helicity window.
Denoting beam current as $I$, the scattered flux is $F = D/I$.
For each quartet $i$, consisting of helicity windows $k=1-4$ of
8.23 msec duration, the raw electron 
cross section asymmetry $A^\mathrm{raw}$
in each HRS is computed from the $F_k$ 
\begin{eqnarray}
  A_i^\mathrm{raw} &=& \hskip 0.05in {\rm sign}_1 \times \left(\frac{F_1-F_2-F_3+F_4}{\sum_k F_k}\right)~,\label{eq:Araw_quad}
\end{eqnarray}
where ${\rm sign}_1 = \pm$ is the sign of the 
first window in the helicity 
pattern $(+--+)$ or $(-++-)$.  
In addition to quadruplets at 120 Hz, the other patterns tried were octets
at 240 Hz and pairs at 60 Hz.  
As the frequency increases, one expects the uncancelled portion
of 60 Hz line $\sigma_{\rm line}$ to contribute more to the noise, while the 
statistical width $\sigma_{\rm stat}$ increases.
Ideally, $\sigma_{\rm line} \ll \sigma_{\rm stat}$ in a window.
Higher frequency also costs statistics because for each helicity flip
there is a fixed deadtime (typ. 0.5 msec) to wait for the helicity electronics to settle.
The 120 Hz flip rate was chosen to be a reasonable tradeoff.
If the beam current and other beam parameters such as position and energy
are stable, and if the electronics noise is well suppressed,
then the statistical uncertainty is dominated by $\sigma_{\rm stat}$
via the detector signals $D$.

Random fluctuations in beam position and energy contributed the 
largest source of noise beyond counting statistics in $A_{raw}$.
For PREX-I, typical beam jitter in window-quadruplets was 
less than 700 parts per million (ppm) in intensity, 
2 parts per million (ppm) in energy, and 20 $\mu$m in position.
The intensity noise was removed through normalization 
to the measured beam intensity, while
noise from the other beam parameters 
was reduced by measuring window differences $\Delta x_i$ using 
beam position monitors and applying a correction $A_{beam}=\sum c_i\Delta x_i$.
This formula is used both to remove stochastic noise from
the beam and to evaluate helicity-correlated corrections.
The $c_i$'s were measured several times each hour
from calibration data in which the beam was modulated
by using steering  coils and an accelerating cavity.
During PREX-I, the 
largest of the $c_i$'s was $\sim$ 50 ppm/$\mu$m.
Details of the corrections can be found 
in refs \cite{KiadThesis,RupeshThesis,LuisThesis,MindyThesis,ZafarThesis}.
The sensitivities for CREX will be smaller because the
cross section for $^{48}$Ca drops less rapidly with
angle than for $^{208}$Pb.

\begin{figure}
\resizebox{0.52\textwidth}{!}{%
\includegraphics{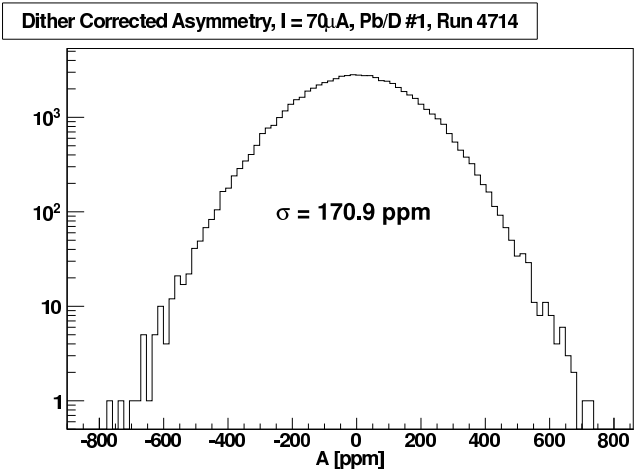}
}
\caption{
Distribution of the asymmetries for a typical PREX-I run at 70$\mu$A.  
Beam-related noise has been subtracted using the 
standard ``dither correction'' method.  The width of
171 ppm is approximately consistent with counting statistics.}
\label{quadasy}
\end{figure}

For PREX-I, the noise in the 
resulting $A_{corr}=A_{raw}-A_{beam}$ was 210~(180)~ppm
per quadruplet, for a beam current of 50~(70)~$\mu$A, dominated by counting statistics ($\sim$ 1~GHz at 70 $\mu$A), see fig ~\ref{quadasy}.

The physics asymmetry $A_{PV}$ is formed from $A_{corr}$ by correcting for the
beam polarization $P_b$ and background fractions $f_i$ with asymmetries $A_i$

\begin{equation}
\label{pvcorr}
A_{PV} = \frac{1}{P_b}\frac{A_{corr} - 
P_b\sum_{i} A_{i}f_{i}}{1-\sum_{i} f_{i}}.
\end{equation}

The PREX-I corrections are shown in 
table~\ref{table:Acorrections}.  The corrections are for
charge normalization, beam asymmetries, the $^{12}$C backing
on the lead target, detector nonlinearities, transverse asymmetries,
and beam polarization.  The total systematic uncertainty was 2.1\% and met
the goals of the experiment.

To compare data to theory, we require a spectrometer acceptance 
function $\epsilon(\theta)$ which characterizes 
the probability, as a function of scattering angle $\theta$, for an electron to 
reach the detector after elastically scattering from $^{208}Pb$.
For example, the asymmetry averaged over the acceptance would be
\begin{equation}
\langle A \rangle = \frac{ \int  d\theta  \sin \theta \hskip 0.02in  A(\theta) \hskip 0.02in \frac{d\sigma}{d\Omega} \epsilon(\theta)}
{\int  d\theta \sin \theta \frac{d\sigma}{d\Omega} \epsilon(\theta)}
\label{eq:accept}
\end{equation}
where $\frac{d\sigma}{d\Omega}$ is the cross section.

Using tracking data, the observed distribution of events corrected
for the cross section, backgrounds, and the effects of 
multiple scattering is used to extract $\epsilon(\theta)$.
To compare the experimental asymmetry to predictions, 
one must integrate the theoretical 
asymmetry over $\epsilon(\theta)$.

\subsection{Transverse Asymmetries}
\label{asy_trans}

A routine and mandatory part of a parity violation experiment 
is to spend about a day measuring
the parity-conserving transverse asymmetry $A_T$ in order to
constrain the systematic error from a possible small transverse 
component of the beam polarization.
The measurement of the $A_T$ itself provides an interesting 
challenge for theoretical prediction, requiring calculation 
of box diagrams with intermediate excited states~\cite{gorchtein,ref:ATdata}.

For these ancillary measurements, the beam polarization is 
set normal to 
the nominal electron scattering plane and the 
asymmetry follows an azimuthal modulation
\begin{equation}
A_T = A_n \vec{P} \cdot \hat{k}
\end{equation}
where $A_T$ is the transverse asymmetry, $A_n$ is the amplitude of the
asymmetry modulation, $\vec{P}$ is the
polarization vector of the electron, and $\hat{k}$ is the unit vector of the
cross product between the incoming and outgoing electron momentum vectors.
This asymmetry is a direct probe to the imaginary part of the
multiple-photon exchange
as it vanishes in the Born-approximation by time reversal symmetry.
The importance of understanding two-photon exchange has 
been highlighted by the discrepancy between $G_E^p$ measurements using 
Rosenbluth-separation and polarization observables~\cite{perdrisat},
attributable to the real part of the multiple-photon exchange amplitude.

Theoretical predictions are challenging to calculate due to the contributions
from hadronic intermediate states in $\gamma-\gamma$ box diagrams and Coulomb
distortion effects which are present for large $Z$.  However, predictions have 
been made that these are on the order of a few ppm with beam energies of 1-2
$\mathrm{GeV}$ and $\theta_e \sim$ 
few degrees using the optical theorem with 
photoabsorption data~\cite{gorchtein} to describe the
intermediate states.  Different approaches, such as using generalized parton 
distributions to describe $e-p$ data~\cite{afanasev}, 
have also been taken.  

\begin{figure}
\resizebox{0.48\textwidth}{!}{%
\includegraphics{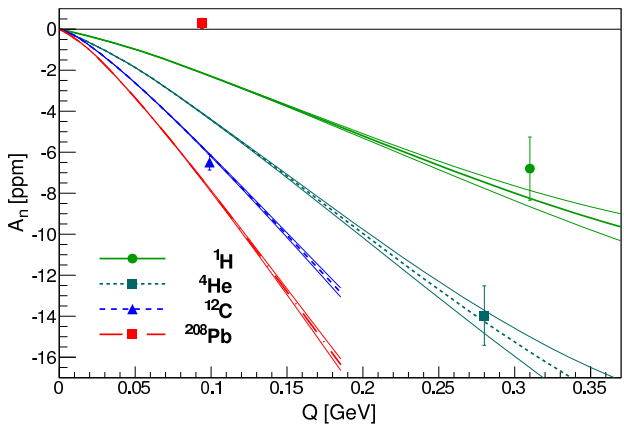}
}
\caption{
Extracted transverse asymmetries $A_n$ vs. $Q$ for several 
different nuclei~\cite{ref:ATdata}.}
\label{fig:at}
\end{figure}

Data for these asymmetries with ${}^1$H, ${}^4$He, ${}^{12}$C, and ${}^{208}$Pb
have been published ~\cite{ref:ATdata}
and are shown in Fig.~\ref{fig:at}.  
There is significant disagreement from theory in ${}^{208}$Pb, the sources of
which are not presently well understood and motivate more
measurements at intermediate $Z$, as well as new calculations
that involve simultaneously Coulomb distortions and dispersion corrections.
In light of this motivation, the CREX experiment measurements
on ${}^{48}$Ca could be useful to help elucidate the dependence 
of these asymmetries on $Z$ and $Q^2$ by providing an additional data point.
Because this asymmetry is so small, directly measuring it requires PV-type 
precision for which this experiment is designed.  A precision of 
$\sim0.5~\mathrm{ppm}$ would be on similar grounds as the previous data
and could be performed in about 1 day.

\subsection{Beam Polarimetry}
\label{polarimetry}

Since the beam polarization $P_b$ is a normalizing factor in
the asymmetry, it must be measured with a high precision
($\frac{dP_b}{P_b} \leq 1$\%).
Developments in 
beam polarimetry are of vital importance to the experimental program
at Jefferson Lab and are ongoing research projects in themselves.
Online monitoring is possible using a Compton
polarimeter which is cross-calibrated using 
M{\o}ller and Mott polarimeters.
Combining the results of the two polarimeters in Hall~A 
we were able to achieve a better
than 1\% accuracy in beam polarization during PREX-I \cite{ref:prexI}
and expect incremental improvements in the uncertainties in the next few years.

In recent years, significant upgrades have been performed for the polarimeters.
The Hall A M{\o}ller polarimeter was upgraded 
with a stronger magnetic field (3T) to ensure a high polarization
of the target foil \cite{hallcpol}.  In addition, the detectors were
segmented and the DAQ upgraded to accommodate higher rates with
lower deadtime.
The Compton polarimeter was also upgraded prior to PREX-I, in order
to achieve an improved figure of merit at low energies by using a 
new green laser and resonant cavity \cite{H3pol,AbdurahimThesis,MeganThesis}. 
Using a new DAQ which integrates the signals from back-scattered photons
we eliminated the systematic error 
from thresholds that affected the older counting method.
For PREX-I, the total systematic uncertainty 
contribution from polarimetry 
totaled 1.2\%, a major accomplishment for 1 GeV running.
At the 2.2 GeV beam energy of CREX, the Compton Polarimeter 
will operate with a higher statistical figure-of-merit and increased resolution of the 
scattered photon spectrum.  
The Compton polarimeter results for the HAPPEX-III experiment~\cite{H3pol},
with a relative systematic error of 0.9\% at 3.4~GeV, are a guide for expected
systematic errors during CREX.  For HAPPEX-III, the systematic error was 
dominated by a 0.8\% uncertainty
in laser polarization.  New techniques for the control of this uncertainty 
have been developed during the Qweak experiment \cite{qweakexpt}.
These will be applied in Hall A 
and can be expected to reduce the photon 
polarization uncertainty to the level of 0.2\%.

\section{PREX-I Result and PREX-II Motivation}
\label{sec.prex}

The ``Lead Radius Experiment'' PREX first ran in 2010 (PREX-I)
and demonstrated successful control of systematic errors,
overcoming many technical challenges, but 
encountered significant loss of beam time due to 
difficulties with vacuum degradation of the target region due
to the high radiation environment \cite{ref:prexI}.
PREX-II is an approved experiment for a followup measurement
with anticipated improvements to take data at a rate
equivalent to the original proposal estimates \cite{ref:prexII}.
PREX measures the parity-violating asymmetry $A_{PV}$ for 
1.06~GeV electrons scattered by about five degrees from $^{208}$Pb.
A major achievement of PREX-I, despite downtimes mentioned above, was
control of the systematic error in $A_{PV}$ at the 2\% level, see
table ~\ref{table:Acorrections} and eq \ref{pvcorr} and the 
discussion below it. 

The result from PREX-I was \cite{ref:prexI}
\begin{equation}
A_{PV}=0.656 \pm 0.060 ({\rm stat}) \pm 0.014 ({\rm syst})\ {\rm ppm}\, .
\label{A.PREXI}
\end{equation}

\begin{figure}
\resizebox{0.52\textwidth}{!}{%
\includegraphics{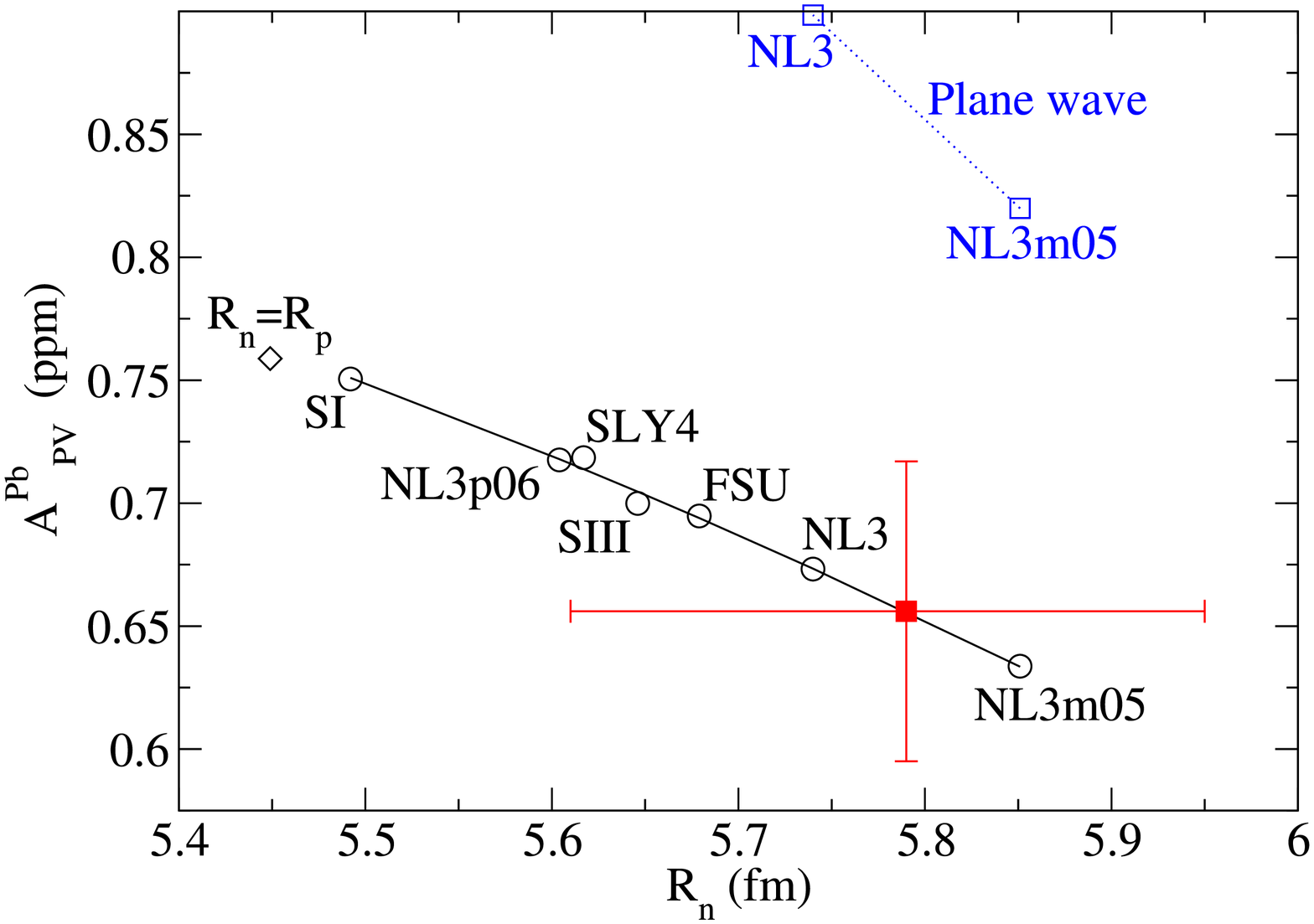}
}
\caption{
Result of the PREX-I experiment (red square) vs neutron point 
radius $R_n$ in ${}^{208}$Pb.  
Distorted-wave calculations for 
seven mean-field neutron densities are 
circles while the diamond marks the expectation for 
$R_n=R_p$ ~\cite{Ban:2010wx}.
\hskip 0.05in References: NL3m05, NL3, and NL3p06 from \cite{ref:nl3},    
FSU from \cite{ref:fsu}, SIII from \cite{ref:siii},
SLY4 from \cite{ref:sly4}, SI from \cite{ref:si}.
The blue squares show plane wave impulse approximation results.}
\label{fig:R}
\end{figure}

This result is displayed in Figure~\ref{fig:R}, in which models predicting the 
point-neutron radius illustrate the correlation of $A_{\rm PV}^{Pb}$ 
and $R_n$~\cite{Ban:2010wx}. 
For this figure, seven non-relativistic and relativistic mean field models 
\cite{ref:nl3,ref:fsu,ref:siii,ref:sly4,ref:si}
were chosen that have charge densities and binding energies in good agreement 
with experiment, and that span a large range in $R_n$.  The weak charge density $\rho_w$ was calculated from model point proton 
$\rho_p$ and neutron $\rho_n$ densities, $\rho_w(r)=q_p\rho_{ch}(r)+q_n\int d^3r'[G_E^p\rho_n+G_E^n\rho_p]$,
using proton $q_p=0.0721$ and neutron $q_n=-0.9878$ weak charges that include radiative corrections.  
Here $G_E^p$ ($G_E^n$) is the Fourier transform of the proton (neutron) electric form factor.  The Dirac equation 
was solved \cite{couldist} for an electron scattering from $\rho_w$ and the experimental $\rho_{ch}$ \cite{chargeden}, 
and the resulting $A_{PV}(\theta)$ integrated over the acceptance, Eq. \ref{eq:accept}, to yield the open circles in Fig. \ref{fig:R}.
The importance of Coulomb distortions is emphasized by indicating results from plane-wave calculations, which are not 
all contained within the vertical axis range of the figure.  

From Eq.~\ref{A.PREXI}, a number of physical quantities were 
deduced~\cite{ref:prexI,ref:prexFF}.  
The form factor $F_W(q)$ of the weak charge 
density $\rho_W(r)$ for $^{208}$Pb is (see eq \ref{F(q)})
\begin{equation}
F_W(q = 0.475~\mathrm{fm^{-1}}) = 0.204 \pm 0.028 .
\end{equation}
Here the total weak charge of $^{208}$Pb is $Q_W$ and $q$ is the momentum 
transfer of the experiment.  
The weak radius of $^{208}$Pb (RMS radius of $\rho_W(r)$) is
 \begin{equation}
 R_W=5.826 \pm 0.181 ({\rm exp}) \pm 0.027 ({\rm mod})\ {\rm fm}.
 \label{RWresult}
 \end{equation} 
 Here the experimental error includes both statistical and systematic 
 effects while the small model error includes model uncertainties 
 related to the surface thickness.  One needs to make very modest 
 assumptions about the surface thickness in order to extract the RMS 
 radius from a single measurement at the particular $Q^2$ chosen for the
 experiment.  Comparing Eq.~\ref{RWresult} to the well-measured charge 
 radius $R_{ch}=5.503$ fm yields a ``weak charge skin''
\begin{equation}
  R_W-R_{ch}=0.323 \pm 0.181 ({\rm exp}) \pm 0.027 ({\rm mod}) \ {\rm fm}.
 \label{weakskin}
 \end{equation}
 Thus the surface region of $^{208}$Pb is relatively enhanced in weak charges
 compared to electromagnetic charges.   This weak charge skin is closely related
 to the expected neutron skin, as discussed below.  Equation~\ref{weakskin}, 
 itself, represents an experimental milestone.  We now have direct evidence 
 that the weak charge density of a heavy nucleus is more extended than the 
 electromagnetic charge density.   Finally the neutron skin thickness, 
the difference of 
 the point neutron $R_n^{208}$ 
and proton $R_p^{208}$  radii of $^{208}$Pb, was
 deduced to be
 \begin{equation}
 R_n^{208}-R_p^{208}=0.33^{+0.16}_{-0.18} \ {\rm fm}.
 \label{eq.prexskin}
 \end{equation}
 This is a (1.8$\sigma$) observation of the neutron skin in a heavy nucleus with
  a purely electroweak reaction. 
PREX-II will have a proposed error in $R_n^{208}$
smaller by a factor of three to $\pm 0.06$ fm.  

To illuminate the importance of the measurement of $R_n$ in nuclear matter, 
 we review some of the implications of the proposed PREX-II measurement of 
neutron radius in  $^{208}$Pb.  The correlation between $R_n^{208}$ and the 
radius of a neutron star, $r_{NS}$, has been shown in models to be very 
strong~\cite{rNSvsRn,gandolfi,erler1211}.
In general, a larger $R_n$ implies a stiffer
EOS, with a larger pressure, that correlates to larger $r_{NS}$ \cite{LSB2013}.
Recently there has been great progress in deducing $r_{NS}$ from X-ray
observations. The value of $r_{NS}$ is deduced from the spectrum and intensity
of the X-rays, with model-dependent corrections for the properties of the 
atmosphere of the neutron star.  The state of the art is as follows.
From observations of X-ray bursts from three-ideal neutron stars, 
$\mbox{Ozel {\it et al}.\ \cite{Ozel:2010fw}}$ find $r_{NS}$ is very small, 
near 10~km,  implying that the EOS softens at high density which is suggestive 
of a transition to an exotic phase of QCD.  In contrast,  
$\mbox{Steiner {\it et al}.\ \cite{Steiner:2010fz}}$, using the same three 
neutron stars plus six more, conclude that $r_{NS}$ is 
near 12 km, leading to a prediction that $R_n^{208}-R_p^{208}=0.15 \pm
0.02$~fm.  This implies a stiffer EOS which leaves little room for softening 
due to a phase transition at high density.

The EOS of neutron-rich matter is closely related to the symmetry energy $S$.  
There is an empirical strong correlation between $R_n^{208}$ and the density 
dependence of the
symmetry energy $dS/d\rho$, with $\rho$ as the baryon density, 
often defined as the parameter
$L = 3\rho (dS/d\rho)$. Data from a wide variety of nuclear reactions are 
being used to constrain $S$ and $L$.  For example,
they can be probed in heavy-ion collisions~\cite{isospindif}; 
$L$ has been extracted from isospin diffusion data~\cite{isospindif2} 
using a transport model.

The symmetry energy $S$ is an important parameter when evaluating the composition 
and structure of a neutron star.  A large $S$ at high density would imply 
a large proton fraction,
which would allow the direct Urca process
($n \rightarrow p + e + \bar\nu_e \hskip 0.04in ; \hskip 0.04in p + e \rightarrow n + \nu_e$) ~\cite{URCA}
for rapid neutrino cooling. 
If $R_n^{208}-R_p^{208}$ were
large, it is likely that massive neutron stars would cool quickly by direct Urca.
In addition, the transition density from a solid neutron star crust
to the liquid interior is strongly correlated with $R_n^{208}-R_p^{208}$~\cite{cjhjp_prl}.  


\begin{table}
\caption{ 
PREX-I corrections to $A_{PV}$ and systematic errors.  See eq \ref{pvcorr} 
and the discussion below it. }
\label{table:Acorrections}
\begin{tabular}{|l|rcl|rcl|}\hline
Correction & \multicolumn{3}{c|}{Absolute (ppb)} & 
\multicolumn{3}{c|}{Relative(\%)}\\ 
\hline \hline
Chg Norm. & \, -84.0 & $\pm$ & $1.5$  & \, -12.8  & $\pm$ & $0.2$ \\
Beam Asy   & 39.0 & $\pm$ & $7.2$  & 5.9  & $\pm$ & $1.1$ \\
Target Backing & $-8.8$   & $\pm$ & $2.6$  & -$1.3$  & $\pm$ & $0.4$ \\
Detector Nonlin.     & $0$  & $\pm$ & $7.6$   & $0$     & $\pm$ & $1.2$\\
Transverse Asy & $0$   & $\pm$ & $1.2$  & $0$  & $\pm$ & $0.2$ \\
Polarization $P_b$ &  70.9 & $\pm$ & 8.3  & $10.8$ & $\pm$ & $1.3$\\
\hline \hline
Total &  17.1 &$\pm$ & 13.7   & 2.6  & $\pm$ & $2.1$\%\\
\hline \hline
\end{tabular}
\end{table}

\section{CREX Proposal}
\label{CREX}

The $^{48}$Ca Radius EXperiment (CREX) was recently approved by the
program advisory committee at Jefferson Lab \cite{ref:CREX}.
The experiment plans to measure the parity-violating asymmetry for elastic
scattering from $^{48}$Ca at E = 2.2~GeV and $\theta=4^{\circ}$.
This will provide a measurement of the weak
charge distribution and hence the neutron
density at one value of Q$^2$ = 0.022 (GeV/c)$^2$. 
It will provide an accuracy in the ${}^{48}$Ca neutron radius $R_n^{48}$
equivalent to $\pm$0.02~fm ($\sim0.6$\%).  As discussed below
in sections ~\ref{subsec.densityfunc} and ~\ref{subsec.coupledcluster},
a measurement this precise 
will have a significant impact
on nuclear theory, providing unique experimental input 
to help bridge ab-initio theoretical approaches 
(based on nucleon-nucleon and  three-nucleon forces) and 
the nuclear density functional theory (based on  energy density functionals).
Together with the PREX measurement of $R_n^{208}$, CREX ($R_n^{48}$) 
will provide unique input in such diverse areas such as neutron star structure, 
heavy ion collisions, and atomic parity violation.
A precise measurement on a small nucleus is favorable because it can be
measured at high momentum transfer where the asymmetry is larger 
(for the proposed kinematics, about 2~ppm).
Also, since $^{48}$Ca is neutron-rich it has a relatively large weak charge
and greater sensitivity to $R_n$.  

\subsection{Testing Density Functional Theory}
\label{subsec.densityfunc}

At the heart of nuclear Density Functional Theory Calculations (DFT)
~\cite{dft_bender,witekEpja,Reinhard1308} is an energy density functional whose minimization 
yields the exact ground state energy and density of a nucleus. However, DFT does not provide a 
practical way to compute the functional.  The commonly 
used EDFs are assumed to have a convenient form in terms of local nucleonic 
densities $\rho_p(r)$ and $\rho_n(r)$ 
and associated currents, involving perhaps a dozen free parameters, and these 
parameters are optimized \cite{kortelainen1,kortelainen2} to reproduce many nuclear observables. 
Using basic observables of stable nuclei, such as binding energies and 
charge radii, the optimization accurately constrains how the functional depends on the isoscalar density 
$\rho_0(r)=\rho_p(r)+\rho_n(r)$ and its gradient ${\bf \nabla}\rho_0(r)$.  

\begin{figure}
\resizebox{0.52\textwidth}{!}{%
\includegraphics{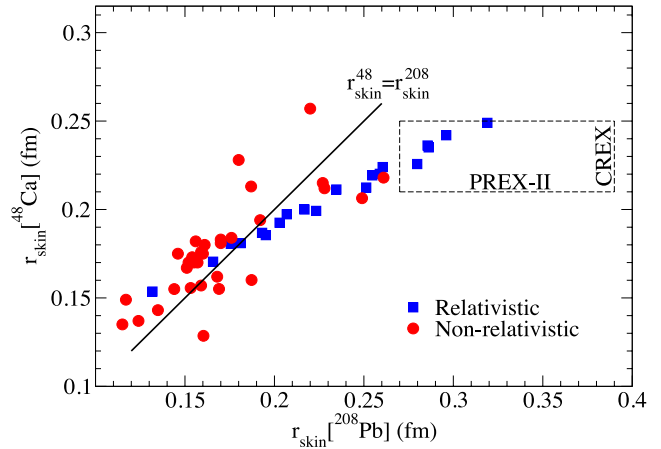}
}
\caption{
Neutron skin thickness in $^{48}$Ca vs skin thickness in $^{208}$Pb showing predictions for several relativistic (blue squares) and non relativistic (red circles) density functionals \cite{Jorgefigure,ref:CREX}.  Also shown is an error box ($\pm$ the expected error bars) for CREX and PREX-II.  This is located at the PREX-I central value for $^{208}$Pb and at an arbitrary $^{48}$Ca skin thickness. }
\label{fig:CavsPb}
\end{figure}

However, there are not many well-measured isovector observables to accurately 
constrain how the functional depends on the isovector density
 $\rho_1(r)=\rho_n(r)-\rho_p(r)$ and ${\bf \nabla}\rho_1(r)$. 
Isovector fields predicted by various 
functionals differ \cite{alphaDJorge,erlernature}; 
hence, the predicted values for the neutron skin vary significantly. 
Remarkably, whereas all the available DFT models predict accurately the binding energy and charge 
radii throughout the nuclear chart,
they are unable to agree on whether $^{48}$Ca or $^{208}$Pb has 
the larger neutron skin \cite{JorgeCREXworkshop}.  In Fig. \ref{fig:CavsPb} we show several relativistic and non relativistic density functional predictions for the neutron skins in $^{48}$Ca and $^{208}$Pb.    These models predict a range of $^{48}$Ca neutron skins that is about seven times larger than the 0.02 fm expected CREX error bar.  In contrast the range in predicted $^{208}$Pb skins is only about 3.5 times the expected 0.06 fm PREX II error bar.  This suggests that CREX may be particularly helpful in constraining density functionals.

The approved PREX-II measurement of $R_{\rm skin}^{208}$, while relevant for astrophysics, 
does not fully constrain the isovector sector of the nuclear density functional. PREX-II is critical in constraining 
the poorly known density dependence
of the symmetry energy, particularly the parameter $L$ that represents the slope of 
the symmetry energy at saturation
density. There is a very strong correlation between $L$ and $R_{\rm skin}^{208}$, so 
at present models with different
values of $L$ predict a large range of neutron skins in $^{208}$Pb, ranging from 
less than 0.1 to greater than 0.3 fm \cite{JorgeCREXworkshop}.
Thus, even the more accurate PREX-II experiment may be 
unable to significantly constraint the
isovector sector of the nuclear density functional.

However, once $L$ is constrained by PREX-II, DFT predicts a correlation
between $R_{\rm skin}^{48}$ and $R_{\rm skin}^{208}$ that is testable with CREX,
see \cite{Reinhard1308}.
For example a large value of $R_{\rm skin}^{208}$ and a small value of $R_{\rm skin}^{48}$ is 
not expected with present EDF parameterizations.  If PREX-II and CREX were to yield such
results, it would strongly suggest that present density functionals incorrectly model 
isovector contributions to the nuclear surface energy (for example gradient terms 
involving $\nabla \rho_1(r)$).  These surface terms are much more important for $^{48}$Ca 
than for $^{208}$Pb because $^{48}$Ca has a larger ratio of surface to volume.  
An additional attractive feature of $^{48}$Ca, as compared to $^{208}$Pb, is 
that the role of electromagnetic effects due to the Coulomb interaction is much reduced in 
the former system, thus allowing a cleaner study of nuclear isovector properties.

We emphasize that PREX-II and CREX together will constrain isovector contributions to the nuclear EDF.  
If PREX-II and CREX results agree with DFT expectations, this provides confidence in theoretical 
predictions of isovector properties  all across 
the periodic table. 
Apart from the inherent importance for nuclear structure physics, these predictions are important 
both for atomic parity experiments and for the extrapolation to very neutron-rich 
systems encountered in astrophysics.  

On the other hand, if PREX-II and CREX results disagree with DFT expectations, 
this will demonstrate that present parameterizations of the isovector part of energy functionals 
are incomplete.  The current  parameterizations are prone to large statistical and systematic   
errors related to isovector terms~\cite{erlernature,Reinhard,Gao}.  
Locating and correcting this error is absolutely essential to develop the 
universal nuclear EDF that will be capable of extrapolating to very neutron-rich 
nuclei and bulk neutron-rich matter. 

\subsection{Ab initio coupled cluster calculations for $^{48}$Ca}
\label{subsec.coupledcluster}

It is important to have a deeper understanding of energy functionals and 
to relate DFT results to underlying 2N and 3N
interactions. Recently there has been considerable progress in ab initio  
coupled cluster calculations for medium mass 
nuclei~\cite{abinitio}.   Hagen~{\it et al.}~\cite{Hagen2012b} have studied neutron rich calcium isotopes with 
large-scale coupled cluster calculations that take advantage of recent computational advances. These calculations 
provide a good description of ground and low lying excited states for a range of calcium isotopes~\cite{CaRadioactive}.

The effects of 3N forces on the neutron density is significant \cite{Holt3N,HebelerFurnstahl,Tews2013}. 
Therefore a measurement 
of $R_{\rm skin}^{48}$ will provide a very useful test of ab initio theory. Present theoretical uncertainties on the
$R_{\rm skin}^{48}$ prediction are large and include contributions from truncating the chiral expansion, the parameters of the 3N force, model space truncations in many body calculations, and omitted terms in the coupled cluster expansion. However the situation is improving rapidly as uncertainty quantification for nuclear structure calculations is an important subject that is receiving considerable attention~\cite{UNEDF,NNpound}.  For example, More~{\it et al.} have developed ways to minimize errors in calculated radii from model space truncations \cite{More2012}. We expect accurate estimations from these ab initio calculations in the near future.  

Note that at this time we do not yet have accurate calculations of $R_{\rm skin}^{48}$ with and without three neutron forces.  We expect the skin to be sensitive to three neutron forces because the pressure of neutron matter was shown to be sensitive to three neutron forces and the neutron skin in $^{208}$Pb is strongly correlated with the pressure of neutron matter.  Therefore it is very important to perform these three neutron force calculations for $^{48}$Ca.

If CREX agrees with the results of coupled cluster calculations this provides a crucial test  of ab initio nuclear structure theory that increases confidence in a variety of nuclear structure predictions and illuminates the role of three-nucleon and in particular three neutron forces.  This is important for a variety of medium mass neutron rich isotopes that are presently being studied with radioactive beams.  It may also be important for calculations of double-beta decay matrix elements.  (The isotope $^{48}$Ca is the lightest nucleus that undergoes double-beta decay and we expect microscopic calculations of double-beta decay matrix elements to be available first for $^{48}$Ca.)  

In contrast, if CREX disagrees with these microscopic calculations, something is likely missing from present ab~initio approaches.  For example, the chiral expansion may not converge as well as hoped because of large $\Delta$ resonance contributions.  This would significantly impact all  nuclear structure theory.

\subsection{CREX Experiment Configuration}

\begin{figure}
\resizebox{0.42\textwidth}{!}{%
\includegraphics{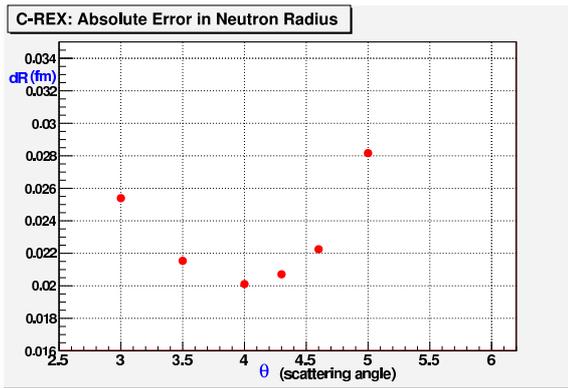}
}
\caption{Error in $R_n$ versus central angle for 2.2 GeV (1-pass beam) for 35 days at 150$\mu$A
for a target thickness of 5\% radiation length.  Maximizing the FOM (see eq \ref{eq:FOM}) 
minimizes the error in $R_n$.   
A 1.2\% systematic error was assumed (see table \ref{table:crex_syserr}) and
added in quadrature to the statistical error.
A total error of 0.02 fm is feasible.  The optimal angle is $4^{\circ}$.}
\label{fig:drr}
\end{figure}

The significant new apparatus elements for CREX
are the ${}^{48}$Ca target and a new $4^{\circ}$ septum magnet. 
The rest of the apparatus is standard equipment and
the methods of section ~\ref{experiment} are applied.
The experiment is designed for $150~\mathrm{\mu A}$
and a 2.2 GeV beam energy, which is a natural beam 
energy at Jefferson Lab (2-passes through the accelerator).
At this energy, the figure-of-merit, which is the total error
in $R_n$ including systematic error, 
optimizes at a scattering angle of $4^{\circ}$, see fig ~\ref{fig:drr}.
Table~\ref{table:Experiments} highlights the experimental 
configuration and goals of PREX and CREX.

\begin{table}\centering
\caption{Parameters of the PREX (I and II) and CREX experiments.  }\begin{tabular}{lccc}
 & PREX & CREX \\ \hline \hline
Energy & 1.0 GeV & 2.2 GeV \\
Angle  & 5 degrees & 4 degrees \\
$A_{PV}$  & 0.6 ppm & 2 ppm \\
$1^{\rm st}$ Ex. State & 2.60 MeV & 3.84 MeV \\
beam current & 70 $\mu$A & 150 $\mu$A \\
rate &  1 GHz & 100 MHz \\
run time & 35 days & 45 days \\
$A_{PV}$ precision  & 9\% (PREX-I) 3\% (PREX-II) & 2.4\% \\ 
Error in $R_N$ &  0.06 fm (PREX-II) & 0.02 fm \\ \hline
\end{tabular}
\label{table:Experiments}
\end{table}

The calcium target will be
a 1 gm/${\rm cm}^2$ isotopically pure $^{48}$Ca target
housed in a vacuum chamber 
with thin entrance and exit windows.
Electrons that scatter from the windows are blocked
(energy-degraded) so that they don't reach the detectors.

The total systematic error goal is 1.2\% on the asymmetry
(see table \ref{table:crex_syserr}) and the
anticipated statistical accuracy is 2.4\%.  
The dominant contributions will be from 
the $Q^2$ determination (see subsection ~\ref{QsqMeas})
and the
polarization measurement (subsection ~\ref{polarimetry}).
These systematic error
contributions are all from effects which have been 
understood and documented by HAPPEX \cite{ref:happex} 
and PREX \cite{ref:prexI},\cite{ref:CREX},
\cite{KiadThesis,RupeshThesis,LuisThesis,MindyThesis,ZafarThesis}.

\begin{table}\centering
\caption[] {Systematic Error Contributions in CREX}
\begin{tabular}{|l|l|}
$Q^2$ & 0.8\% \\
Polarization & 0.8\%\\
Charge Normalization & 0.1\%\\
Beam Asymmetries & 0.3\%\\
Detector Non-linearity & 0.3\%\\
Transverse & $0.1$\%\\
Inelastic Contribution & 0.2\% \\\hline
Total & 1.2\%
\end{tabular}
\label{table:crex_syserr}
\end{table}

\section{Conclusions}
\label{conclusions}

In this paper we discussed the future measurements 
PREX-II and CREX at Jefferson Lab.  
The parity-violating electron
scattering asymmetry from $^{208}$Pb and $^{48}$Ca 
provide a clean measurement at one $Q^2$ of the 
weak charge of these nuclei and 
are sensitive to the nuclear symmetry energy.
The experiments leverage the advantages Jefferson Lab,
with it's highly stable and precisely controlled electron
beam and the high resolution spectrometers, which are 
uniquely suited to perform these experiments.
Within the next few years, these $R_n$ measurements 
on $^{208}$Pb and $^{48}$Ca 
will provide powerful experimental inputs to tune nuclear models 
of increasing sophistication. 

PREX-I achieved the first electroweak observation,
at the 1.8$\sigma$ level, of the neutron skin of
$^{208}$Pb and successfully demonstrated this technique for
measuring neutron densities, with an excellent control of systematic errors.
The future PREX-II run will reduce the uncertainty by a factor of three,
to $\pm 0.06$ fm in $R_n$.
While PREX-II will put a constraint on the density 
dependence of the symmetry energy (the parameter $L$), 
models predicting neutron radii of medium mass and light nuclei 
are affected by nuclear dynamics beyond $L$. 
CREX will provide new and unique input into the isovector sector
of nuclear theories, and the high precision measurement 
of $R_n$ ($\pm 0.02$ fm) in a doubly-magic nucleus 
with 48 nucleons will help build a critical bridge between 
ab-initio approaches and nuclear DFT.   
CREX results can be directly compared to new coupled cluster calculations sensitive to three neutron forces.

The authors gratefully acknowledge all the collaborators on
the PREX-II ~\cite{ref:prexII} and CREX ~\cite{ref:CREX} 
proposals and the participants at the
CREX 2013 workshop \cite{CREXworkshop}, and especially the
discussions with G. Hagen, J. Mammei, D. McNulty, W. Nazarewicz,
K. Paschke, J. Piekarewicz, S. Riordan, and P.A. Souder.
This work was supported by the U.S. Department of Energy, 
grants DE-FG02-88R40415-A018 (University of Massachussets) and
DE-FG02-87ER40365 (Indiana University), and by the
Jefferson Science Associates, LLC,  which operates Jefferson Lab 
for the U.S. DOE under U.S. DOE contract DE-AC05-060R23177.


\end{document}